\documentclass[
preprintnumbers
]{revtex4}
\usepackage{graphicx}

\begin{document}

\preprint{YITP-03-69}

\title{The BABAR resonance as a four-quark meson\footnote{Talk given
  at Hadron2003, the X-th International Conference on Hadron
spectroscopy, August 31 -- September 6, 2003, Aschaffenburg, Germany}
}

\author{K. Terasaki}
\affiliation{Yukawa Institute for Theoretical Physics,
             Kyoto University, Kyoto 606-8502, Japan}

\date{May 20, 2003}


\begin{abstract}

A possible assignment of the new resonance observed at the $B$ 
factories to a four-quark meson, 
$\hat F_I^+ \sim \{[cu][\bar s\bar u] - [cd][\bar s\bar d]\}$, 
is proposed and two-body decays of four-quark mesons through $I$-spin 
conserving strong interactions are studied. It is expected
that some of them can be observed as narrow resonances.  
Implication of existence of four-quark mesons in hadronic weak 
interactions is also discussed. 

\end{abstract}

\maketitle

Recently a scalar $D_s^+\pi^0$ resonance with 
a mass $\simeq 2.32$ GeV and a width $\sim 10$ MeV 
has been observed at the $B$ factories~\cite{B-fact}. 
The above values of the mass and the width have been obtained from 
Gaussian fits, and then it has been concluded that the intrinsic 
width is equal to or narrower than $\sim 10$ MeV. 
While it has been interpleted as an ordinary $c\bar s$ state such as 
the chiral partner of $D_s^+$~\cite{chiral} or the excited $c\bar s$ 
state~\cite{CJ}, their predicted mass values are still in 
controversy~\cite{mass}. 
In addition, possible assignments to exotic hadron states such as a 
($DK$) molecule~\cite{BCL} (or atom~\cite{Szczepaniak}), an isospin 
($I$) singlet four-quark meson~\cite{CH}, a mixed scalar state of a 
$c\bar s$ and an iso-singlet four-quark meson~\cite{BP}, etc. have 
also been proporsed. In all the above proporsals, however, the 
dominant decay of the new resonance into the $D_s^+\pi^0$ final state 
proceeds through $I$-spin non-conserving interactions since it 
has been assigned to an iso-singlet state, and therefore it should be 
extremely narrow. In this talk, we will propose another possible 
assignment to a singly charged component 
of iso-triplet scalar four-quark mesons in contrast with the above
proposals.  

Four-quark mesons, $\{qq\bar q\bar q\}$, can be classified into
four types~\cite{Jaffe}, 
\begin{equation}
\{qq\bar q\bar q\} = [qq][\bar q\bar q] \oplus (qq)(\bar q\bar q) 
\oplus \{[qq](\bar q\bar q)\pm (qq)[\bar q\bar q]\}, 
\label{eq:Jaffe}
\end{equation} 
where parentheses and square brackets denote symmetry and 
anti-symmetry, respectively. 
Only the $[qq][\bar q\bar q]$ and $(qq)(\bar q\bar q)$ can have 
$J^{P(C)}=0^{+(+)}$ at the lowest level. Each of them is again 
classified into two classes since there exist two different 
configurations to produce color singlet scalar states, i.e., 
${\bf 6}\times{\bf \bar 6}$ and ${\bf \bar 3}\times{\bf 3}$ of color 
$SU_c(3)$. However, these two can mix with each other, so that 
they are classified into heavier and lighter ones. The former is 
dominated by the ${\bf 6}\times{\bf \bar 6}$ of $SU_c(3)$ while the 
latter by the ${\bf \bar 3}\times{\bf 3}$ since the force between two 
quarks is repulsive when they form a ${\bf 6}$ of $SU_c(3)$ while 
attractive when ${\bf \bar 3}$~\cite{Hori}. We discriminate these two 
by putting $\ast$ on the former in accordance 
with Ref.~\cite{Jaffe} in which the four-quark mesons were studied
within the framework of $q=u,\,d$ and $s$. It can be extended 
straightforwardly to the one including $q=u,\,d,\,s$ and 
$c$~\cite{charm}. 

In this talk, we concentrate on the $[cq][\bar q\bar q]$ mesons 
(with $q=u,\,d,\,s$). We now assign the new resonance to 
$\hat F_I^+ \sim \{[cu][\bar s\bar u] - [cd][\bar s\bar d]\}$ 
and estimate the masses of the lighter class of the 
$[cq][\bar q\bar q]$ mesons using the simple quark counting and 
taking 
$\Delta m_s = m_{s} - m_n \simeq 0.1$ GeV, ($n=u,\,d$),  
at $\sim 2$ GeV scale and the measured $m_{\hat F_I} = 2.32$ GeV as 
the input data~\cite{Terasaki-PRD03}. To estimate the mass values of 
the heavier class of $[cq][\bar q\bar q]$ mesons, we use 
$\Delta m_c = m_c - m_n \simeq 1.5$ GeV and 
$m_{\hat \kappa^*}\simeq 1.6$ GeV~\cite{Jaffe} 
as the input data.  Their estimated mass values are listed in
{\bf TABLE~1}. 

\begin{center}
\begin{quote}
{\bf TABLE~1.} Ideally mixed scalar $[cq][\bar q\bar q]$ mesons 
(with $q=u,\,d,\,s)$, where 
$S$ and $I$ denote the strangeness and the isospin. 
\end{quote}
\vspace{0.5cm}

\begin{tabular}
{c c c c c}
\hline
$\,\, S\,\,$
&$\,\, I=1\,\,$
&$\,\,I={1\over 2}\,\,$
&$\,\, I=0\,\,$
&$\,\,$Mass(GeV)$\,\,$
\\
\hline
$1$
&
\begin{tabular}{c}
$\hat F_I$ \\
$\hat F_I^*$
\end{tabular}
&
&
\begin{tabular}{c}
$\hat F_0$\\
$\hat F_0^*$
\end{tabular}
&
\begin{tabular}{c}
{\hspace{3mm}2.32($\dagger$)}\\
{(3.1)}
\end{tabular}
\\
\hline
0
&
&
\begin{tabular}{c}
{$\hat D$}\\
{$\hat D^*$}\\
{$\hat D^s$}\\
{$\hat D^{s*}$}
\end{tabular}
&
&\begin{tabular}{c}
{2.22}\\
{(3.0)}\\
{2.42}\\
{(3.2)}
\end{tabular}

\\
\hline 
-1
&
&
&\begin{tabular}{c}
$\hat E^0$\\
$\hat E^{0*}$
\end{tabular}
&\begin{tabular}{c}
{2.32}\\
(3.1)
\end{tabular}
\\
\hline 
\end{tabular}\vspace{2mm}\\
\hspace{-30mm}
$(\dagger)$ : Input data
\end{center}

As seen in {\bf TABLE~1}, the four-quark mesons with $\ast$ have 
large masses enough to decay into two vector mesons in addition to 
two pseudoscalar mesons, so that they will be broad. On the contrary, 
the estimated masses of $[cq][\bar q\bar q]$ without $\ast$ are near 
(or lower than) the thresholds of two body decays through $I$-spin 
conserving strong interactions, so that their phase space volumes are 
small even if kinematically allowed. Besides, it is
seen~\cite{yitpws}, from the crossing matrices for color and spin in 
Ref.~\cite{Jaffe}, that the wavefunction overlapping between the 
scalar $[qq][\bar q\bar q]$ meson and the two pseudoscalar $q\bar q$ 
meson state is small since the former is dominated by the 
${\bf \bar 3}\times{\bf 3}$ of color $SU_c(3)$ and the 
${\bf 1}\times{\bf 1}$ of spin $SU(2)$. Therefore, some of them can 
decay through $I$-spin conserving interactions but their rates will 
be small and they can be observed as narrow resonances such as the 
new resonance. Since some of them are not massive enough to decay 
into two pseudoscalar mesons through $I$-spin conserving 
interactions, their dominant decays may be $I$-spin non-conserving 
ones (unless their masses are higher than the expected ones). 

The $\hat F_I$ mesons form an iso-triplet, $\hat F_I^{++}$, 
$\hat F_I^+$ and $\hat F_I^0$, where the $I$-spin symmetry is always 
assumed in this talk unless we note. Then all of them can have the
same type of kinematically allowed decays, 
$\hat F_I\rightarrow D^+_s\pi$, with different charge states. 
The $\hat D$ and $\hat D^s$ form two independent iso-doublets. 
The $\hat D$ can decay into $D\pi$ final states and the kinematical 
condition is similar to the one in the decay, 
$\hat F_I\rightarrow D^+_s\pi$, as long as the mass value of $\hat D$ 
in {\bf TABLE~1} is taken. The dominant decay of $\hat D^s$ which 
contains an $s\bar s$ pair would be 
$\hat D^s\rightarrow D\eta^s \rightarrow  D\eta$. 
Because of $m_{\hat D^s}\simeq m_D + m_\eta$, however, it is not 
clear if such a decay is allowed kinematically, as long as the value 
of $m_{\hat D^s}$ in {\bf TABLE~1} is taken. Even if allowed, the 
rate would be much smaller than the ones for the above decays because 
of smaller phase space volume. 
The $\hat F_0^+$ is an iso-singlet counterpart of the $\hat F_I$
mesons. It cannot decay into the $D^+_s\pi^0$ as long as the $I$-spin
is conserved, so that it will decay dominantly through $I$-spin 
non-conserving interactions. In this case, the width of the 
$\hat F_0^+$ will be extremely narrow (much narrower than the 
$\hat F_I$ and $\hat D$ mesons). If its mass should be higher (by 
$\sim 50$ MeV or more) because of some $I$-spin dependent force, it 
could decay dominantly into the $DK$ final states (but could be still 
narrow.) 
The $\hat E^0$ is an iso-singlet scalar meson with charm $C=1$ and 
strangeness $S=-1$, i.e., $\hat E^0 \sim [cs][\bar u\bar d]$. 
It cannot decay into $D\bar K$ final states unless it is massive 
enough. If its mass is almost the same as the $\hat F_0^+$, it cannot 
decay through strong interactions or electromagnetic 
interactions~\cite{exotic} as there are no ordinary mesons with $C=1$ 
and $S=-1$. If it can be created, therefore, it will have a very long 
life. 

Now we study numerically decays of the $[cq][\bar q\bar q]$ mesons. 
Consider a decay, 
$A({\bf p})\rightarrow B({\bf p'})\, +\, \pi({\bf q})$,   
as an example. 
The rate for the decay is given by 
\begin{equation}
\Gamma(A \rightarrow B\pi)
= {1\over 2J_A + 1}{q_c\over 8\pi m_A^2}
\sum_{spins}|M(A \rightarrow B\pi)|^2,
                                                    \label{eq:rate}
\end{equation} 
where $J_A$, $q_c$ and $M(A \rightarrow B\pi)$ denote the spin of $A$, 
the center-of-mass momentum of the final mesons and the decay 
amplitude, respectively. To calculate the amplitude, we here use the 
PCAC (partially conserved axial-vector current) hypothesis and a hard 
pion approximation in the infinite momentum frame (IMF), i.e., 
${\bf p}\rightarrow\infty$~\cite{hard-pion,suppl}. In this 
approximation, the amplitude is evaluated at a slightly unphysical 
point, i.e., $m_\pi^2\rightarrow 0$, and is given approximately by 
the asymptotic matrix element of $A_\pi$, 
$\langle{B|A_{\bar \pi}|A}\rangle$, 
as 
\begin{equation}
M(A \rightarrow B\pi) 
\simeq \Bigl({m_A^2 - m_B^2 \over f_\pi}\Bigr)
                              \langle{B|A_{\bar \pi}|A}\rangle,
                                                     \label{eq:amp}
\end{equation} 
where $A_{\pi}$ is the axial counterpart of the $I$-spin. 
{\it Asymptotic matrix elements} (matrix elements taken between 
single hadron states with infinite momentum) of $A_{\pi}$ can be 
parameterized by using {\it asymptotic flavor symmetry} 
(flavor symmetry of the asymptotic matrix elements). For asymptotic 
symmetry and its fruitful results, see Ref.~\cite{suppl} and
references therein. We here list only the related asymptotic matrix 
elements~\cite{charm}, 
\begin{eqnarray}
&&\langle{D_s^+|A_{\pi^-}|\hat F_I^{++}}\rangle 
= \sqrt{2}\langle{D_s^+|A_{\pi^0}|\hat F_I^{+}}\rangle 
= \langle{D_s^+|A_{\pi^+}|\hat F_I^{0}}\rangle
 \nonumber\\
&&=-\langle{D^0|A_{\pi^-}|\hat D^{+}}\rangle            
= 2\langle{D^+|A_{\pi^0}|\hat D^{+}}\rangle 
            \nonumber\\
&&=-2\langle{D^0|A_{\pi^0}|\hat D^{0}}\rangle 
= -\langle{D^+|A_{\pi^+}|\hat D^{0}}\rangle. 
                                                 \label{eq:axial-ch}
\end{eqnarray} 
Inserting Eq.(\ref{eq:amp}) with Eq.(\ref{eq:axial-ch}) into 
Eq.(\ref{eq:rate}),  we can calculate approximate rates for the 
allowed two-body decays mentioned before. Here we equate the 
calculated width for the $\hat F_I^+\rightarrow D_s^+\pi^0$ decay 
to the measured one of the new resonance, i.e., 
$\Gamma(\hat F_I^+\rightarrow D^+_s\pi^0) \sim 10\,\, {\rm MeV}$,   
as an example, since we do not find any other decays which can have 
large rates, and use it as the input data when we estimate the rates 
for the other decays. [However, the number $\sim 10$ MeV should not
be taken too literally since it is still tentative, i.e., a 
possibility to take 
$\Gamma(\hat F_I^+\rightarrow D^+_s\pi^0) \sim {a\,\, few\,\, 
{\rm or}\,\, several}$  MeV 
is not excluded.] The results are listed in {\bf TABLE~2}. All the 
calculated rates of the $\hat F_I$ and $\hat D$ mesons are lying in 
the region near the input data, so that they will be observed as 
narrow resonances in the $D_s^+\pi$ and $D\pi$ channels,
respectively. 
As for the $\hat D^s\rightarrow D\eta$ decays, because of 
$m_{\hat D^s}\simeq m_D\,+\,m_\eta$, it is not clear if they are 
kinematically allowed. Besides, the decay is sensitive to the 
$\eta$-$\eta'$ mixing scheme which is still model 
dependent~\cite{Feldmann}. Therefore, we 
\begin{center}
\begin{quote}{
{\bf TABLE~2.} Dominant decays of scalar $[cq][\bar q\bar q]$ mesons 
and their estimated widths. 
$\Gamma(\hat F_I^+\rightarrow D_s^+\pi^0) \sim 10$ MeV is used as the 
input data. 
The decays into the final states between angular brackets are not 
allowed kinematically as long as the parent mass values in the 
parentheses are taken.
}

\end{quote}
\vspace{0.5cm}

\begin{tabular}
{c c c }
\hline
\begin{tabular}{c}
Parent \\
(Mass in GeV)
\end{tabular}
&Final State
&

Width 
(MeV)
\\
\hline
\begin{tabular}{c}
$\hat F_I^{++}(2.32)$ \\
$\hat F_I^+(2.32)$\\
$\hat F_I^0(2.32)$
\end{tabular}
&
\begin{tabular}{c}
$D_s^+\pi^+$\\
$D_s^+\pi^0$\\
$D_s^+\pi^-$
\end{tabular}
& $\sim$10
\\
\hline
{$\hat D^+(2.22)$}
&\begin{tabular}{c}
{$D^0\pi^+$}\\
{$D^+\pi^0$}
\end{tabular}
&
\begin{tabular}{c}
{$\sim$10}\\
{$\sim$ 5}
\end{tabular}
\\
{$\hat D^0(2.22)$}
&\begin{tabular}{c}
{$D^+\pi^-$}\\
{$D^0\pi^0$}
\end{tabular}
&
\begin{tabular}{c}
$\sim$10\\
$\sim$ 5
\end{tabular}
\\
\hline 
{$\hat D^s(2.42)$}
&$D\eta$ 
&-- 
\\
\hline 
{$\hat F_0^+(2.32)$}
&\begin{tabular}{c}
{$<D_s^+\eta>$}\\
$D_s^+\pi^0$
\end{tabular}
&\begin{tabular}{c}
--\\
($I$-spin viol.)
\end{tabular}
\\
\hline 
{$\hat E^0(2.32)$}
&
$<D\bar K>$
&
--
\\
\hline 
\end{tabular}\vspace{2mm}\\
\end{center}
need more precise and 
reliable values of $m_{\hat D^s}$, $\eta$-$\eta'$ mixing parameters 
and decay constants in the $\eta$-$\eta'$ system to obtain a definite 
result. 

In summary we have studied the decays of the scalar 
$[cq][\bar q\bar q]$ mesons into two pseudoscalar mesons by assigning 
the new resonance to the $\hat F_I^+$ and assuming the $I$-spin 
conservation. All the allowed decays are not very far from the 
corresponding thresholds, so that their rates have been expected to 
saturate approximately their total widths. Therefore, we have used 
the measured width as the input data.  
The $\hat F_I$ and $\hat D$ could be observed as narrow resonances 
such as the new observation. It is very much different from the 
results in Ref.~\cite{CH} in which the new resonance was assigned to 
$\tilde D_{0s}$ [$\hat F_0^+$ in our notation] and all the other 
$[c\bar qq\bar q]$ mesons were predicted to be much broader 
($\sim 100$ MeV or more). To distinguish the present assignment from
the other models and to confirm it, therefore, it is important to 
observe these narrow resonances. 
Although we have not studied numerically, 
we can qualitatively expect that the $\hat D^s$ will be much narrower 
than the $\hat F_I$ and $\hat D$ mesons. 
The $\hat F_0^+$ decays through $I$-spin non-conserving interactions,
so that it should be extremely narrow. 
The $\hat E^0$ will decay through weak interactions if it is created 
as long as its mass is below the $\hat E^0\rightarrow D\bar K$
threshold. 

We have studied, so far, the strong decay properties of a group of 
the four-quark mesons. If their existence is confirmed, it will be 
very helpful in understanding of hadronic weak decays of $K$ and 
charm mesons. The heavier class of $[qq][\bar q\bar q]$ and 
$(qq)(\bar q\bar q)$ mesons can play an important role in hadronic
weak decays of charm mesons, since the masses of some of the related 
members are expected to be close to the ones of the parent charm
mesons. For example, the expected mass, 
$m_{\hat\sigma^{s*}}\simeq 1.8$ GeV,  of the 
$\hat \sigma^{s*}\sim [us][\bar s\bar u]+[ds][\bar s\bar d]$ 
is close to the $m_{D^0}$ but the one, 
$m_{\hat\sigma^*}\simeq 1.45$ GeV, of the 
$\hat\sigma^* \sim [ud][\bar d\bar u]$ 
is much lower than the $m_{D^0}$ as seen in Ref.~\cite{Jaffe}.  
Therefore, the former can contribute to the intermediate state of the 
$D^0\rightarrow K^+K^-$ decay and enhance strongly the decay while 
the latter can contribute to the $D^0\rightarrow \pi^+\pi^-$ decay 
but cannot so strongly enhance it. In this way, we can obtain a 
solution to the long standing puzzle~\cite{PDG02},  
\begin{eqnarray}
&&{\Gamma(D^0\rightarrow K^+K^-)
                     \over \Gamma(D^0\rightarrow \pi^+\pi^-)}
\simeq 3,
\end{eqnarray}
consistently with the other two-body decays of charm 
mesons~\cite{charm}. 
Furthermore, the lighter $(qq)(\bar q\bar q)$ mesons are useful to 
understand the violation of the $|\Delta {\bf I}|= 1/2$ rule 
in $K\rightarrow \pi\pi$ decays consistently with the $K_L$-$K_S$ 
mass difference, the $K_L\rightarrow \gamma\gamma$ and the Dalitz 
decays of $K_L$~\cite{Terasaki01}. 

Confirmation of the existence of four-quark mesons is very much
important not only in hadron spectroscopy but also in hadronic weak
interactions of $K$ and charm mesons.


\section*{Acknowledgments}

The author would like to thank Professor T.~Onogi for providing 
information of the new resonance, discussions and encouragements, 
and Professor T.~Kunihiro for discussions and encouragements.

\end{document}